\documentclass[11pt,a4paper]{article}

\usepackage{memoraxpaper}
\usepackage{longtable}
\usepackage[
  backend=biber,
  style=authoryear-comp,
  maxcitenames=2,
  maxbibnames=12,
  uniquename=init,
  giveninits=true,
  doi=true,
  url=true,
  isbn=false,
  eprint=true
]{biblatex}
\DeclareNameWrapperFormat{labelname}{%
  \ifcitation{\bibhyperref{#1}}{#1}}
\AtBeginBibliography{\raggedright\sloppy}

\reporttitle{EngramBench: A Capability-Grounded Benchmark for Skill-Evolution Harnesses}
\reportrunningtitle{EngramBench}
\reportauthors{Zhixuan Tan\textsuperscript{1,*}, Pengjie Gu\textsuperscript{2,*}, Zhao Li\textsuperscript{2}, Yihan Hu\textsuperscript{2}, Xu He\textsuperscript{2}, Dong Li\textsuperscript{2}, Jianye Hao\textsuperscript{2}}
\reportaffiliations{\textsuperscript{1}The Chinese University of Hong Kong, Shenzhen\\\textsuperscript{2}MemoraX AI}
\reportversion{V 1.0}
\reportdate{Sep 2026}
\reportcontact{%
  \textsuperscript{*}Equal contribution.\\
  Corresponding author:
  \href{mailto:jianye.hao@tju.edu.cn}{jianye.hao@tju.edu.cn}\\
  Code: \url{https://github.com/Virgil-Tan/EngramBench}%
}
\reportfooter{MemoraX AI Research · EngramBench}
\reportlogo{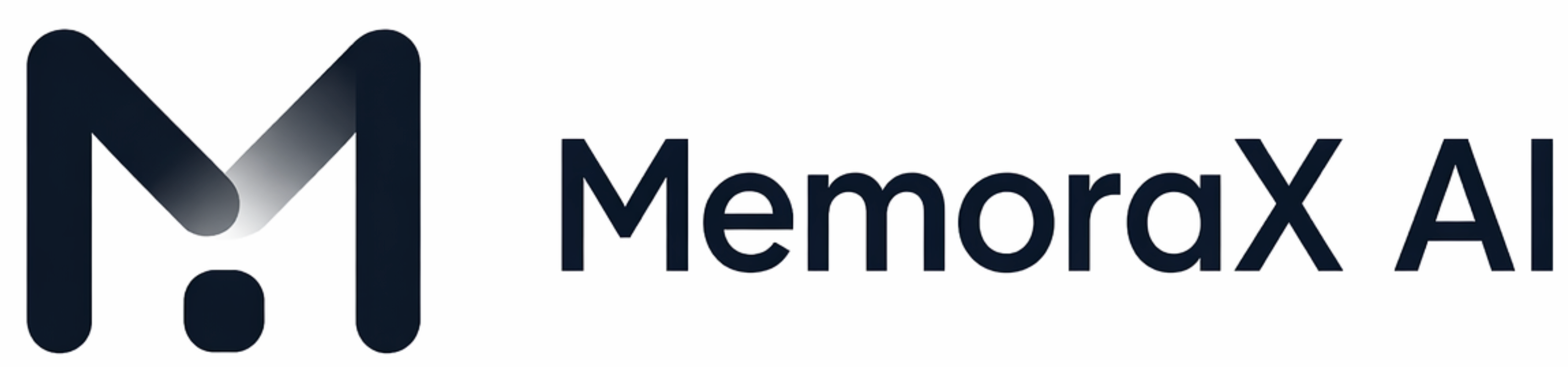}
\reportsubject{EngramBench: A Capability-Grounded Benchmark for Skill-Evolution Harnesses}
\reportkeywords{long-horizon software engineering, agent benchmarks, skill evolution, procedural memory, cross-project transfer}
\begin{document}

\makememoraxpapertitle

\begin{reportabstract}
While large language models have achieved remarkable success in isolated code generation, authentic software engineering requires sustained reasoning, complex state management, and continuous cross-domain abstraction. However, current evaluations of skill evolution in autonomous agents suffer from a critical identifiability problem: they structurally confound genuine capability abstraction with rote solution leakage (i.e., copying highly similar code from historical training data). To resolve this, we introduce \textbf{EngramBench}, a rigorous, capability-grounded benchmark governed by the strict axiom of \textit{capability overlap without solution overlap}. Comprising 30 diverse learning tasks and 13 unseen transfer tasks, EngramBench challenges agents to navigate interactive, multi-hour development cycles driven by LLM-simulated users. Our extensive evaluation across 48 multi-hour execution trajectories—corroborated by human-expert validation—reveals a profound insight into procedural memory. We demonstrate that static skill banks do not magically bypass the ``last mile'' of exact code implementation, which remains bottlenecked by the base model's inherent reasoning limits. However, they serve as an indispensable \textit{execution compass}. By navigating agents away from catastrophic, token-heavy trial-and-error, genuine capability abstraction slashes redundant context bloat and reduces overall coding time by over 55\%. Ultimately, EngramBench shifts the evaluation paradigm from trivial pattern matching to the verifiable measurement of deep, cross-domain capability transfer.
\end{reportabstract}

\printreportkeywords

\section{Introduction}
\label{sec:intro}
Large Language Models (LLMs) have demonstrated strong capabilities in code generation and problem-solving \parencite{chen2021codex,yang2024sweagent}. Building on this foundation, the field is increasingly focusing on \textit{skill evolution} \parencite{wang2023voyager,li2026skillhone}. By equipping agents with memory and reflection mechanisms, this approach allows them to accumulate engineering experience and learn from execution feedback over time \parencite{shinn2023reflexion}. Rather than treating every task as an isolated challenge, agents can learn from past successes and mistakes, evolving from static models into lifelong learners. Real-world software engineering requires agents to distill reusable practices from past projects and apply them to new, unfamiliar contexts. Therefore, establishing a benchmark to measure this cross-project learning is critical for determining whether AI agents can operate as continuous software engineers rather than one-off script generators. 

However, evaluating skill-learning frameworks poses a fundamental design challenge: structurally decoupling genuine skill abstraction from rote code memorization. This challenge is driven by a task-design dilemma. On the one hand, constructing learning and testing tasks within the same environment (e.g., a single repository) can ensure knowledge relevance, but reusable procedures must be separated from task-specific answer paths to prevent ``solution leakage'' \parencite{skillsbench2026}. Without this boundary, agents can achieve artificially high scores by simply copying historical abstract syntax trees (ASTs) or specific APIs, masquerading memorization as learning. Conversely, preventing solution reuse is not enough: learning and testing tasks must retain transferable skill dependencies for historical experience to carry instructional value \parencite{zheng2025lifelongagentbench}. Without both safeguards, evaluations risk either rewarding solution duplication or providing irrelevant histories, producing diagnostic ambiguity. Ultimately, when an agent fails on a new project, researchers cannot isolate the root cause: did the memory framework fail to abstract the skill, or did the benchmark simply provide the wrong history?

This ambiguity exposes a fundamental flaw in current evaluations: the \textit{confusion problem}. Downstream scores confound two entirely separate variables:
\begin{equation}
\resizebox{0.9\linewidth}{!}{$\displaystyle \textbf{Downstream Score} = f\Big(\underbrace{\text{Distillation Capability}}_{\text{\scriptsize \textit{What we aim to measure}}} \times \underbrace{\text{Presence of Evidence in History}}_{\text{\scriptsize \textit{What the benchmark must guarantee}}}\Big)$}
\end{equation}
By treating the right side as an unverified black box, existing benchmarks make the left side impossible to isolate. A high score might merely reflect the base model's zero-shot intuition, while a low score might simply mean the required knowledge was absent from the historical tasks. To escape this trap and properly isolate the \textit{distillation capability}, we must look to human developers. Human learning is a \textit{many-to-many} abstraction process, not a 1-to-1 code mapping. An engineer who learns "transaction rollbacks" in e-commerce and "state synchronization" in IoT does not copy specific APIs when building a new medical system. The concrete \textbf{solutions} are entirely different, but the abstracted engineering \textbf{capabilities} are reused. Real learning happens when core challenges overlap, but specific implementations do not. Thus, a scientific benchmark must strictly enforce a golden design rule:
\begin{equation}
\textbf{Capability Overlap Without Solution Overlap}
\end{equation}
By strictly ensuring tasks share high-level challenges, we make the \textit{presence of evidence} required in our first equation structurally auditable. Crucially, this axiom clearly delineates the boundaries of experience transfer: it strictly forbids the retention of project-specific implementations, hardcoded business constants, or exact API paths (solution leakage), while explicitly expecting the abstraction of general transaction templates, fault-recovery pseudocode, and cross-cutting engineering workflows (capability overlap). By running them on completely isolated systems with zero shared codebase, we physically prevent cheating. This axiom finally breaks the design trap and enables the measurement of true skill abstraction.

\begin{figure}[!t]
\centering
\includegraphics[width=\linewidth]{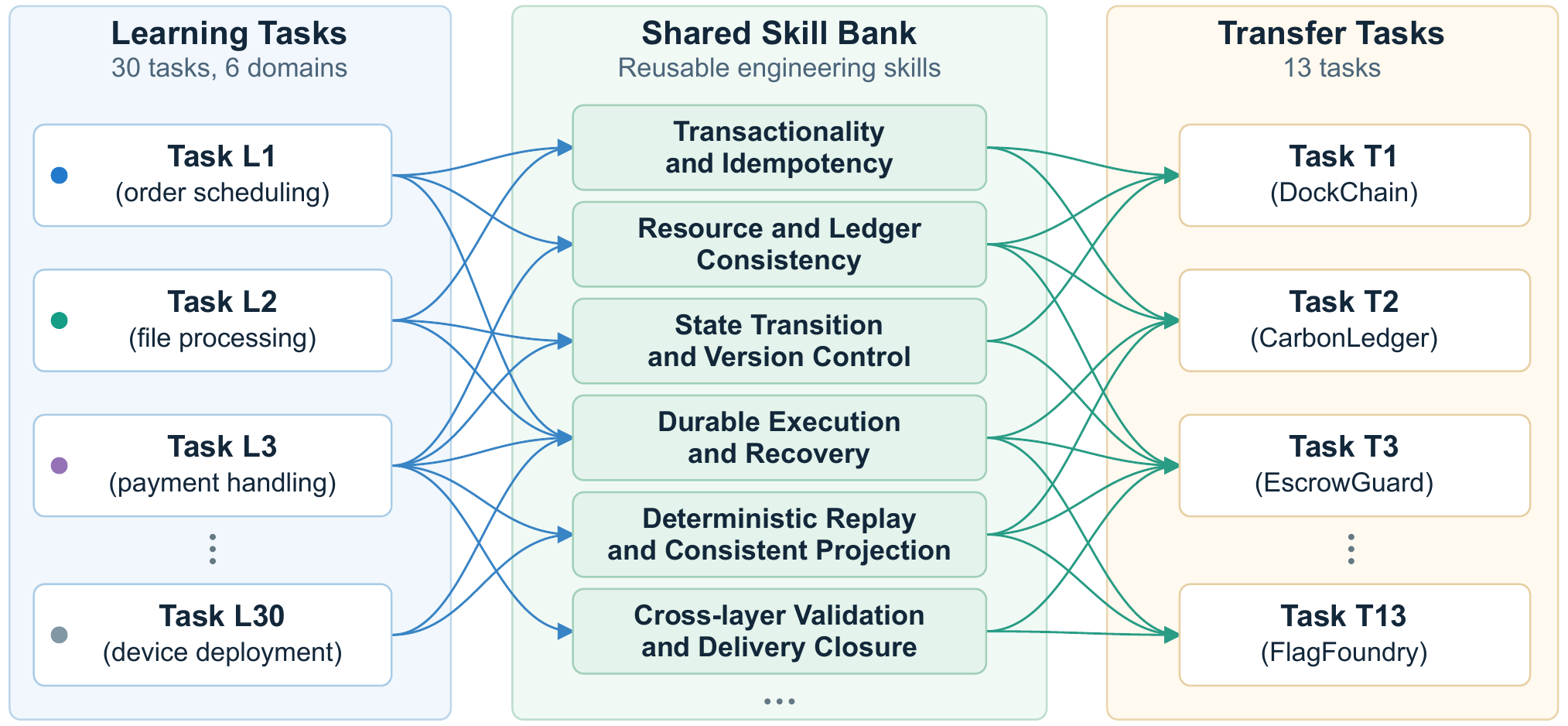}
\caption{Many-to-many capability transfer in EngramBench. Experience from different learning tasks is abstracted into reusable skills that support multiple transfer tasks. Arrows denote illustrative capability relationships, not direct reuse of project-specific code.}
\label{fig:many-to-many}
\end{figure}

Figure~\ref{fig:many-to-many} illustrates this many-to-many transfer of reusable capabilities across otherwise distinct projects.

To materialize this design axiom, we introduce \textbf{EngramBench}, a \textit{strictly decoupled}, capability-grounded benchmark built to evaluate true skill evolution. First, to guarantee the \textbf{presence of learning evidence}, EngramBench features 30 representative learning tasks across 6 diverse business domains. Rather than merely providing the final correct code, the benchmark supplies \textit{feedback-rich trajectories}---including the agent's initial mistakes, test-case failures, and iterative fixes. This ensures the framework has sufficient data to abstract underlying engineering principles. Second, to enforce \textbf{solution isolation} and evaluate \textbf{many-to-many composition}, we design 13 long-horizon transfer projects. These test projects share zero business logic or APIs with the learning tasks. To succeed, the agent cannot simply retrieve old code; it must dynamically combine abstract skills learned from different domains. Crucially, to test sustained execution beyond static code generation, our transfer projects are not single-turn tests. Instead, they are multi-turn, interactive software projects driven by an \textbf{LLM-simulated user}, following an interaction paradigm established in tool-agent-user evaluation \parencite{taubench2024}. Throughout the multi-hour development cycle, this simulated user actively asks follow-up questions, checks progress against the fixed requirements, and rejects incomplete drafts. Reliable multi-turn execution presents a distinct challenge beyond single-turn capability \parencite{laban2025lost}. EngramBench tests whether distilled, long-term skills help agents meet that challenge. Finally, to maintain strict scientific discipline, EngramBench enforces a \textit{two-phase decoupled protocol}. Once the learning phase is complete, the agent's generated skill bank is strictly frozen, enforcing a strict information boundary between historical distillation and transfer execution. For controlled evaluation, we standardize the base model and learning trajectories across all methods, allowing us to purely compare the algorithmic efficiency of different skill-distillation frameworks.

To demonstrate the value of this rigorous design, our evaluations establish a clear baseline for future research. We show that while current base models naturally struggle with these long-horizon tasks without prior knowledge, a proper skill-evolution framework can produce a real, positive \textit{learning delta}, successfully guiding the agent through complex, multi-turn user interactions.

In summary, the main contributions of this work are as follows:
\begin{itemize}
    \item \textbf{A Clear Rule for Skill Evolution:} We clearly define the \textit{identifiability problem} in current benchmarks and introduce a strict design rule: \textit{capability overlap without solution overlap}.
    
    \item \textbf{Building EngramBench:} We introduce a \textit{strictly decoupled, capability-grounded} benchmark. It includes 30 diverse learning tasks with rich feedback, and 13 unseen, complex testing tasks driven by multi-turn simulated users.
    
    \item \textbf{Strict Evaluation Protocols:} We design a two-phase frozen testing process. With these fair and strict rules, we provide the community with a reliable standard to measure true skill evolution without hidden bias.
\end{itemize}

\Needspace{4\baselineskip}
\section{Related Work}
\label{sec:related}

\subsection{Code Generation vs. Software Engineering}
The evaluation of Large Language Models (LLMs) in programming has historically centered on short-horizon code generation. Benchmarks such as HumanEval~\parencite{chen2021codex} and MBPP~\parencite{austin2021program} evaluate the ability of models to synthesize standalone, algorithmic functions from natural language prompts. While these benchmarks effectively measure a model's static knowledge of syntax and basic algorithms, real-world software engineering extends far beyond isolated code snippets~\parencite{jimenez2023swe}. Authentic software engineering is a long-horizon, dynamic process characterized by massive contextual dependencies, environmental feedback, and iterative debugging across multiple modules. To succeed in these environments, autonomous agents cannot rely solely on static pre-training; they must possess the ability to continuously interact, learn from execution failures, and accumulate reusable experience. This fundamental shift---from single-turn code generation to continuous software engineering---necessitates new paradigms for evaluating how agents abstract and reuse engineering skills over time. Recent empirical work by METR evaluates task-completion reliability as a function of task length, measured by the time human experts require, highlighting the need to assess sustained execution beyond short tasks \parencite{metr2025measuring}. EngramBench directly addresses this need through long, interactive development lifecycles.

\subsection{Current Agent Benchmarks and Their Limitations}
To evaluate agents in realistic development environments, the community has introduced repository-level benchmarks, most notably SWE-bench~\parencite{jimenez2023swe} and its variants~\parencite{swebenchlite2024}. These benchmarks represent a significant milestone by challenging agents to resolve real-world GitHub issues within complex, real-world codebases. Recently, some evaluations have attempted to adapt these static benchmarks into lifelong learning or skill-evolution scenarios by providing agents with historical issue-resolution trajectories as training data. 

However, when repurposed for skill evolution, these benchmarks suffer from the structural dilemma outlined in Section~\ref{sec:intro}. Because their task sequences are inherently intra-domain---typically requiring the agent to solve consecutive issues within the exact same repository (e.g., \textit{Django} or \textit{SymPy})---they inadvertently trigger \textit{solution leakage}. Agents can achieve artificially high transfer scores simply by retrieving and copying highly similar abstract syntax trees (ASTs), specific local APIs, or database schemas from the historical tasks. Consequently, these benchmarks test an agent's capacity for exact code retrieval and rote context-matching, rather than its ability to distill cross-cutting engineering capabilities. EngramBench diverges fundamentally from these paradigms by explicitly enforcing cross-domain task sequences, ensuring capability overlap without solution overlap.

\subsection{Procedural Memory and Skill Evolution in LLMs}
To bridge the gap between static LLMs and lifelong learning agents, recent literature has explored architectures equipped with procedural memory or skill libraries. Frameworks such as Voyager~\parencite{wang2023voyager}, ExpeL~\parencite{zhao2023expel}, and various agentic continual learning systems~\parencite{agentcl2026} attempt to distill execution trajectories into reusable text guidelines or executable code repositories. These harnesses aim to abstract past successes and failures into a persistent memory bank, which is then retrieved to guide decision-making in novel environments. 

Despite the algorithmic innovations introduced by these harnesses, their reported efficacies remain confounded by the benchmarks on which they are evaluated. Because existing evaluations fail to decouple a method's true abstraction capability from the sheer availability of highly similar in-domain data, the true utility of current procedural memory mechanisms remains opaque. We cannot ascertain whether an agent has genuinely learned an abstract principle (e.g., concurrent state synchronization) or merely memorized a repository-specific workaround. EngramBench provides the missing, capability-grounded infrastructure necessary to causally verify and benchmark these procedural memory frameworks.

\section{Constructing EngramBench}
\label{sec:construction}

Unlike traditional benchmarks that repurpose isolated GitHub issues for static code completion, EngramBench is architected as a requirement-driven, long-horizon software engineering sandbox. It challenges agents to synthesize complete business systems from lightweight API contracts, forcing them to navigate cross-layer dependencies, environmental constraints, and interactive feedback.

\subsection{Task Curation \& Domains}
Instead of scraping disparate open-source commits, EngramBench features meticulously constructed, requirement-driven sandbox projects that mirror real-world business constraints. The accumulation phase consists of 30 learning tasks strategically distributed across 6 core business domains: \textit{Resource Booking \& Logistics, Data \& Collaboration, Transaction \& Ledger, IoT \& Configuration, Security \& Moderation, and Infrastructure \& Communication}. This taxonomy guarantees that abstract engineering capabilities (e.g., transactional idempotency, failure recovery) appear repeatedly across disjoint business contexts.

\begin{figure}[!htbp]
\centering
\includegraphics[width=\linewidth]{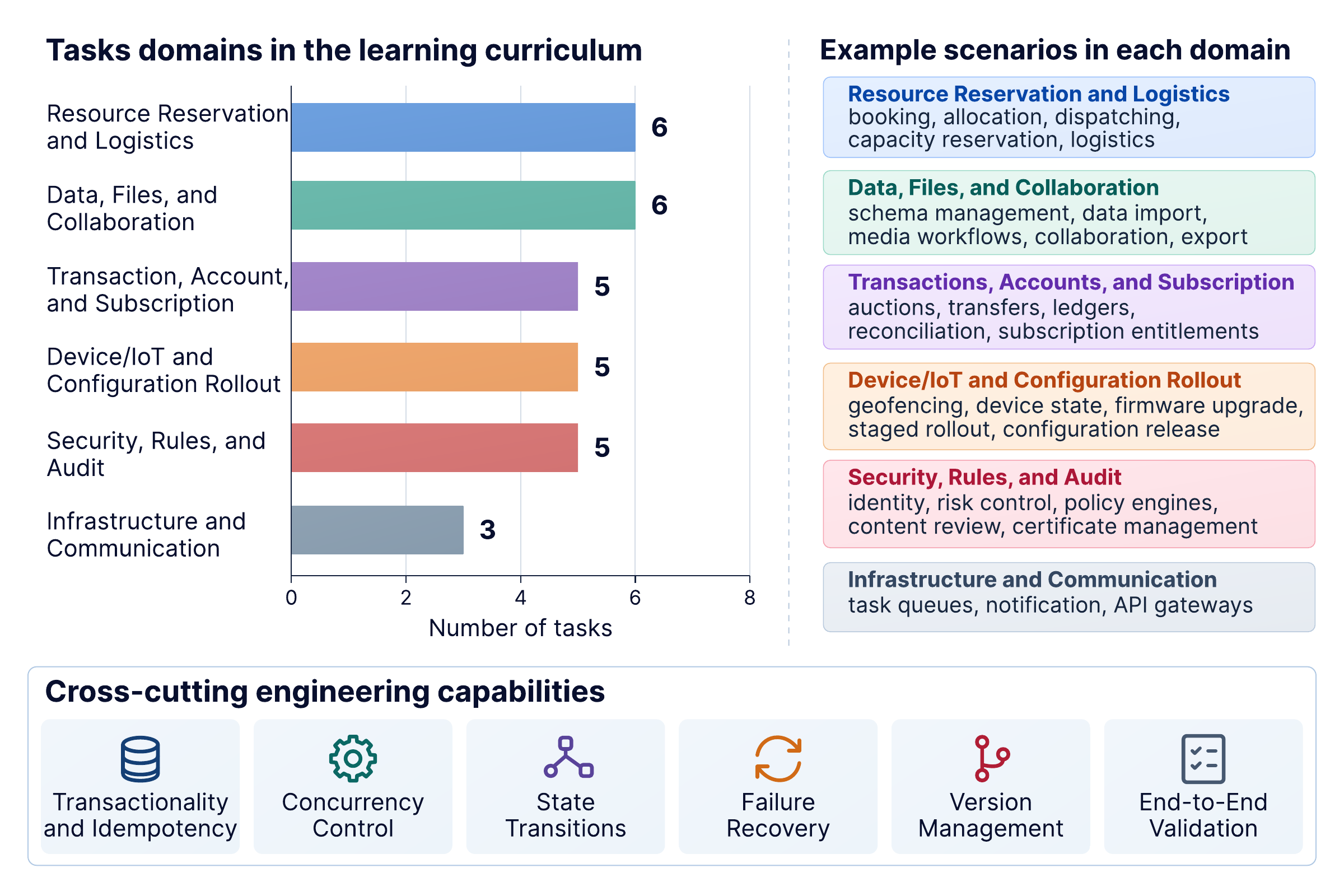}
\caption{The 30 learning tasks span six business domains. Cross-cutting engineering capabilities recur across domains, providing a basis for skill abstraction and transfer.}
\label{fig:task-domains}
\end{figure}

Figure~\ref{fig:task-domains} summarizes the learning-task distribution and the cross-cutting capabilities shared across domains.

Each task follows a strict developmental lifecycle: a definitive README contract, a frozen execution plan, and a scenario-driven progression. Agents are provided with a deterministic scaffolding that defines request/response contracts, initialization hooks, and public interface checkers. To provide a rigorous capability-grounded foundation, the learning phase yields 30 complete, feedback-rich development trajectories. The current dataset archives 583 interactive execution turns, capturing simulated user dialogue, agent decisions, tool invocation events, compiler outputs, and hidden evaluation results. Crucially, all tasks and distilled skills have been rigorously verified through human-expert baseline executions to ensure their feasibility and intrinsic engineering utility.

Complementing the learning tasks, the benchmark includes a transfer evaluation set of 13 tasks. These tasks share no business logic with the accumulation set but are explicitly designed to test the cross-domain extraction of the underlying engineering capabilities mapped in the learning curriculum.

To ground this capability overlap in actual learning experiences, we make the evidence directly inspectable. We systematically map representative public requirements from the transfer tasks to reusable engineering mechanisms, tracing them back to concrete implementation, repair, and verification episodes in the learning histories. These links are established through recorded execution trajectories---such as local failure-repair loops or fault-injection validations---rather than superficial similarities in task names. Table~\ref{tab:evidence_mapping} presents representative mappings that validate this capability coverage. Appendix~\ref{app:evidence-mapping} provides the complete task-level mapping across all 13 transfer tasks.

\begin{table}[htbp]
\centering
\caption{Learning Evidence Mapping for Representative Transfer Requirements}
\label{tab:evidence_mapping}
\small
\setlength{\tabcolsep}{3.5pt}
\renewcommand{\arraystretch}{1.12}
\begin{tabularx}{\linewidth}{@{}>{\raggedright\arraybackslash}p{0.15\linewidth}>{\raggedright\arraybackslash}p{0.20\linewidth}>{\raggedright\arraybackslash}X>{\raggedright\arraybackslash}p{0.14\linewidth}l@{}}
\toprule
\textbf{Transfer Task} & \textbf{Reusable Mechanism} & \textbf{Learning Episode (Source)} & \textbf{Evidence Type} & \textbf{Eval} \\
\midrule
\textit{CreatorRights\-Exchange}
& Append-only correction and read-projection consistency
& \textit{BillForge} (T7): Ledger reads omitted adjustment entries; a query repair restored their visibility.
& Local repair loop
& B-10 \\
\addlinespace
\textit{IncidentRelay}
& Stable event identity and persistent retry handling
& \textit{QueueForge} (T13): SIGKILL before ACK persistence. \textit{ConfigRelay} (T12): HTTP 503 followed by same-event retry and ordering checks.
& Impl + Fault injection
& C-07 \\
\addlinespace
\textit{CarbonLedger}
& Publish consistency during post-generation crash
& \textit{ArtifactVault} (T26): SIGKILL injected after file rename but before DB commit.
& Impl + Verification
& C-03 \\
\addlinespace
\textit{CapacityLease}
& Expired lease recovery and backlog drain throughput
& \textit{QueueForge} (T16): Insufficient recovery count; worker batching and test-counting fixes followed by a passing re-test.
& Local repair loop
& E-06 \\
\bottomrule
\end{tabularx}

\par\smallskip
{\footnotesize
Note: \textit{Evidence Type} distinguishes implementation-and-verification sequences from iterative local failure-repair loops recorded in execution trajectories. T denotes the recorded turn number; ArtifactVault T26 is from the original development history, while the other cited turns are from V2 migration histories. These are mechanism-level links, not proof that entire transfer tasks were solved in the learning phase. The full task-level table appears in Appendix~\ref{app:evidence-mapping}.
}
\end{table}

\subsection{Codebase Complexity and Cross-Layer Implementation}
EngramBench operates in a modern, full-stack engineering environment primarily utilizing TypeScript, Node.js 22, React, and PostgreSQL 16. Rather than dropping agents into massive legacy codebases to perform localized one-line bug fixes, the benchmark evaluates the ability to build complex, cross-layer implementations from a unified structural foundation. 

For the 13 transfer tasks, the initial execution environment provides a clean V2 API scaffold. On average, each starter repository contains 23.46 files (including documentation and contract definitions) and 8 source code files, totaling 622 non-empty lines of code (LOC) and 268 public operations across the entire transfer set. 

The true complexity of EngramBench lies in its deep engineering constraints rather than raw initial LOC. Agents are required to implement full-stack business logic that spans database migrations, background dispatchers, OpenAPI integrations, and browser-observable UI states. For instance, in a task like \textit{CapacityLease}, the agent must not only implement a booking endpoint but also guarantee time-range capacity conservation, atomic admission control, concurrent race-condition handling, and crash recovery. These challenges demand that the agent orchestrates persistent state synchronization and fault tolerance, decisively elevating the evaluation beyond superficial syntax generation.

\subsection{Environment Capabilities and Multi-Layered Feedback}
To support these long-horizon tasks, EngramBench provides an isolated, stateful OCI-compliant (Docker) execution sandbox. The environment natively provisions necessary toolchains, including shell access, file I/O operations, source code search utilities, TypeScript/npm build environments, PostgreSQL databases, and Chromium for end-to-end (E2E) browser testing. 

A critical differentiator of EngramBench is its dynamic, multi-layered feedback loop, which organically guides the agent through multi-hour development cycles:
\begin{itemize}
    \item \textbf{Execution Feedback:} Agents autonomously invoke tools and immediately receive real-world standard output/error, including compiler traces, database connection states, and unit test failures.
    \item \textbf{Interactive Progression:} A Simulated User Agent actively monitors the execution against the frozen plan. This multi-turn interaction paradigm has precedent in benchmarks such as $\tau$-bench, which pairs a language agent with a simulated user and domain-specific tools \parencite{taubench2024}. Rather than injecting new out-of-scope requirements, the simulated user queries progress, handles blockages, and forces the coding agent to cross-reference the original README to determine missing implementations.
    \item \textbf{Delivery and Evaluation:} The development cycle is governed by a strict two-stage delivery pipeline. Agents first iterate against a public interface checker. Once the public contracts are satisfied, the code is strictly frozen and dispatched to an isolated hidden evaluation environment. Here, comprehensive testing---spanning integration, concurrency, performance, and crash-recovery scenarios---generates the definitive test feedback, determining whether the underlying engineering capabilities were successfully synthesized.
\end{itemize}

\paragraph{The Long-Horizon Scale.} 
Unlike traditional benchmarks that execute in a single prompt-completion cycle, EngramBench accurately simulates the marathon nature of real-world software engineering. A single transfer task typically demands 25 to 30 interactive development turns, characterized by continuous exchanges between the Simulated User and the Coding Agent. Throughout this prolonged process, the agent must continuously maintain context, parse massive execution logs, and navigate cascading errors, culminating in hours of continuous execution and tens of millions of processed tokens per project.

\section{Rigorous Evaluation Protocol}
\label{sec:evaluation}

To ensure that performance gains genuinely reflect capability abstraction rather than mere context matching or test-set overfitting, EngramBench establishes a highly stringent, multi-dimensional evaluation protocol governed by physical isolation boundaries.

\subsection{The Confirmed-Case Pass Fraction (CCPF)}
Traditional code generation benchmarks predominantly rely on binary Pass@1 metrics. However, for long-horizon software engineering, binary success is overly coarse: it fails to distinguish between an agent that implemented 95\% of the core business logic but missed an edge case, and an agent that failed to even initialize the database. 

To provide a high-resolution signal of requirement coverage, we introduce the \textit{Confirmed-Case Pass Fraction} (CCPF). For a given task $t$ with an applicable hidden test case set $\mathcal{C}_t$, the CCPF is defined as:
\begin{equation}
\mathrm{CCPF}_t = \frac{1}{|\mathcal{C}_t|} \sum_{c\in\mathcal{C}_t} \mathbf{1}[c\text{ passes}] = \frac{P_t}{|\mathcal{C}_t|}
\end{equation}
where $P_t$ represents the number of confirmed passed cases, and $\mathcal{C}_t$ is the set of all applicable, non-excluded evaluation cases. All applicable cases are equally weighted, and the metric evaluates the finalized delivery without arbitrarily excluding challenging edge cases. Crucially, self-authored unit tests generated by the agent are explicitly ignored. The denominator $|\mathcal{C}_t|$ strictly includes all applicable evaluation cases. Consequently, unresolved states, compilation crashes, or unreached test cases remain in the denominator but contribute nothing to the passed numerator, implicitly penalizing incomplete executions. For macro-level performance across multiple runs, we utilize the micro-average CCPF to accurately reflect the holistic completion rate of complex engineering contracts.

\subsection{Multi-Dimensional Verification and Hidden Stress Testing}
EngramBench evaluates systems far beyond superficial API responses. As demonstrated by the EvalPlus framework \parencite{evalplus2023}, inadequate test suites can fail to expose incorrect code and distort model rankings. To reduce this ``fake-green'' risk, our hidden evaluators systematically stress-test the agent's implementation across five critical engineering dimensions: (A) \textit{Contracts \& Lifecycle}, (B) \textit{Data Idempotency \& Concurrency}, (C) \textit{Worker Recovery \& Durability}, (D) \textit{Cross-Layer UI Consistency}, and (E) \textit{Operational Compatibility}.

The hidden evaluators employ aggressive, adversarial testing mechanisms to verify true engineering capabilities. For example, in the \textit{CapacityLease} task, the evaluator does not simply check the successful return of a booking API. Instead, it enforces:
\begin{itemize}
    \item \textbf{Concurrency \& Idempotency:} The evaluator fires 64 concurrent requests with identical idempotency keys to multiple API instances. It strictly asserts that the database yields exactly one lease and one corresponding business event, penalizing naive implementations that bypass race conditions.
    \item \textbf{Controlled Crash \& Recovery:} The evaluator intentionally pauses a background worker and terminates it via \texttt{SIGKILL}. It then spawns a replacement worker to verify if the uncompleted task is correctly resumed and finished without duplicating business side-effects.
    \item \textbf{Anti-Hallucination (Fake-Green) Checks:} To prevent agents from writing vacuous tests that always pass, the evaluator dynamically breaks dependencies (e.g., swapping the database for an unreachable address or masking the Chromium path). If the agent's integration or E2E tests still pass under these broken conditions, the submission is failed for hallucinating execution logic.
\end{itemize}

\subsection{Anti-Cheating and Sandboxing Protocols}
To eradicate \textit{solution leakage} and ensure causal evaluation, EngramBench enforces a strict \textit{two-phase decoupled protocol} supported by OCI-compliant (Docker) container isolation.

\paragraph{Phase 1: Skill Generation \& Freezing.} 
During the learning phase, the harness distills historical trajectories into a reusable Skill Bank (e.g., \texttt{SKILL.md}). Crucially, this output is compiled into a manifest with SHA-256 hashes. This guarantees deterministic, immutable inputs for the downstream evaluation.

\paragraph{Phase 2: Isolated Transfer Execution.} 
For the 13 transfer tasks, the agent is instantiated in a completely independent environment. It is provided with the current task's README, a clean workspace scaffold, and the \textit{frozen} Skill Bank. The agent is physically restricted from mounting or accessing the raw workspaces, interaction histories, or codebases from the learning phase. Furthermore, skill evolution is strictly disabled during this phase, preventing the agent from continuously optimizing its memory bank using test-set feedback. This freeze applies within each run: the S-Full-20 bank was selected by qualitative coverage of engineering procedures rather than transfer scores, whereas the Coreless-3 ablation was designed after inspecting the preceding S-Full/S-Ablated comparison and is therefore exploratory.

\paragraph{Hidden Evaluation Boundaries.} 
To prevent look-ahead bias, the hidden evaluators are entirely unmounted during the development phase. Once the agent passes the public interface checker, the harness automatically freezes the commit and logs a secure code digest. The evaluation phase then mounts this frozen snapshot into a separate, dedicated testing container. This multi-stage isolation keeps hidden-test code outside the development workspace, reducing opportunities for direct test-specific optimization and supporting evaluation of transfer to new projects.

\section{Experimental Setup}
\label{sec:experimental_setup}

To empirically validate the necessity of procedural memory, we conduct a highly controlled ablation study. We focus on four representative transfer tasks (\textit{CapacityLease}, \textit{EscrowGuard}, \textit{MeterSettle}, and \textit{FlagFoundry}) and evaluate four skill configurations. With three independent trials per configuration-task pair, this yields 48 strictly audited execution trajectories.

Across all runs, the primary Coding Agent is powered by GPT-5.5 Medium. To standardize the interactive progression and eliminate human variability, we deploy DeepSeek V4 Pro (\texttt{thinking/high} mode) as the Simulated User Agent. The developmental boundaries are strictly enforced: the task \texttt{README} defines the immutable requirements, and the \textit{Frozen Plan} dictates the initial architectural strategy. Fixing the plan reduces variation in initial planning and focuses the comparison on \textit{execution capability}; controlled experiments show that even when explicit knowledge and plans are provided, per-step execution accuracy can degrade as the horizon grows \parencite{illusion2025long}. The user \textit{Scenario} strictly drives the interaction forward without injecting out-of-scope business requirements.

\subsection{Baselines: Native Skills vs. Ablations}
To disentangle the value of the provided memory from the base model's inherent reasoning, we deploy the following configurations:

\begin{itemize}
    \item \textbf{No Skills (Zero-Shot Baseline):} Evaluates the model's native capability to navigate long-horizon challenges relying solely on its pre-trained weights.
    
    \item \textbf{S-Full-20 (Complete Memory):} A static repository of 20 consolidated skills generated via a rigorous \textit{two-stage distillation and compression pipeline}. First, the full historical trajectories of the 30 learning tasks are locally distilled into a massive raw skill bank (368 skill packages). Second, a global compression phase (using DeepSeek V4 Pro at temperature 0) semantically merges redundant mechanisms and eliminates generic advice, retaining exactly 20 highly actionable and cross-domain transferable engineering principles.
    
    \item \textbf{S-Ablated-20 (History Filtering):} To test the framework's sensitivity to evidence quality, this bank is derived from a deliberately weakened historical dataset. Targeted removal of critical implementation, debugging, and verification records reduces the raw historical text volume by 69.66\%. Crucially, these deprived trajectories undergo the \textit{exact same} two-stage pipeline and compression prompt as \textit{S-Full-20}, distilling down to 20 skills. This isolates the impact of severe evidence deprivation on the final synthesized knowledge.
    
    \item \textbf{Coreless-3 (Capability Ablation):} A surgically edited version of the \textit{S-Full-20} bank that removes the targeted core software engineering procedures (e.g., concurrency control), leaving only 3 basic interface skills. While this exploratory ablation reduces both the volume and scope of the memory, it serves to test whether the base model can sustain performance when these procedures are absent from the supplied memory.
\end{itemize}

\subsection{Evaluation Metrics}
A robust skill-evolution framework must balance functional correctness with computational efficiency. As emphasized by prior work on agent evaluation \parencite{agentsmatter2024}, success rates alone are insufficient; evaluation should jointly measure accuracy and cost. Accordingly, we report token consumption and development time alongside functional correctness.

\paragraph{1. Functional Correctness.}
We report task success using the \textit{Micro-Average CCPF}, governed by the strict rules defined in Section~\ref{sec:evaluation}. Performance deltas are strictly reported in absolute \textit{percentage points}.

\paragraph{2. Computational Cost (Tokens \& Time).}
We track the exact resource burden introduced by the developmental loop:
\begin{itemize}
    \item \textbf{Total Tokens:} Merges input and output (reasoning) tokens. We strictly deduplicate cached contexts by session to reflect true computational overhead.
    \item \textbf{Pure Development Time:} The Coding Agent's exact wall-clock execution time, explicitly subtracting all off-cycle phases (e.g., Simulated User processing, hidden evaluations, and verified pauses).
\end{itemize}

\paragraph{3. Behavioral Dynamics.}
To differentiate between an agent that "gives up early" and one that "works efficiently", we track internal actions per turn: command execution counts, patch invocations, reasoning token generation, and context compaction triggers. Reductions in resource overhead are reported as \textit{relative percentages} against the \textit{No Skills} baseline.

\section{Experimental Results \& Analysis}
\label{sec:results}

We evaluate four configurations---\textit{No Skills}, \textit{S-Full-20}, \textit{S-Ablated-20}, and \textit{Coreless-3}---across 48 strictly audited execution trajectories. Rather than merely reporting binary success rates, our analysis deconstructs the exact behavioral impact of procedural memory, revealing a profound shift in how agents navigate long-horizon engineering tasks.

\subsection{Aggregate Performance: The ``Last Mile'' Bottleneck}
We first evaluate the impact of skills on functional correctness using the micro-average Confirmed-Case Pass Fraction (CCPF) across the 12 runs for \textit{No Skills} and \textit{S-Full-20}.

\begin{table}[htbp]
\centering
\caption{Task-Level CCPF: No Skills vs. S-Full-20 (12 Runs Each)}
\label{tab:performance_verified}
\small
\begin{tabularx}{\linewidth}{@{}Yrrr@{}}
\toprule
\textbf{Task} & \textbf{No Skills} & \textbf{S-Full-20} & \textbf{Change (pp)} \\
\midrule
MeterSettle & 76.42\% & 80.49\% & +4.07 \\
FlagFoundry & 46.34\% & 47.15\% & +0.81 \\
EscrowGuard & 18.52\% & 33.33\% & +14.81 \\
CapacityLease & 47.83\% & 46.38\% & -1.45 \\
\midrule
\textbf{Micro-Average} & 46.63\% & \textbf{51.25\%} & \textbf{+4.62} \\
\bottomrule
\end{tabularx}
\end{table}

As shown in Table~\ref{tab:performance_verified}, injecting the complete skill bank provides a verifiable overall gain (+4.62 percentage points). However, the improvement is uneven. While complex transactional tasks like \textit{EscrowGuard} see massive gains (+14.81 pp), \textit{CapacityLease} experiences a slight regression.

This non-monotonic benefit exposes a critical philosophical insight into autonomous learning: \textit{a reusable engineering principle is not an executable implementation}. Procedural memory provides a high-level architectural compass (e.g., ``how to structure idempotent transactions''), but it intentionally omits exact syntax. The ``last mile'' of engineering---writing the precise code, bridging specific APIs, and passing edge-case assertions---still fundamentally relies on the base model's inherent reasoning limits. If the base model struggles with final execution constraints, even the best navigational guide cannot single-handedly guarantee a perfect pass rate.

\subsection{The Efficiency Paradigm: Memory as an Execution Compass}
While correctness improvements are bound by the base model's ``last mile'' capabilities, the impact of the skill bank on execution efficiency is transformative. 

\begin{table}[htbp]
\centering
\caption{Average Efficiency Metrics per Run: No Skills vs. S-Full-20}
\label{tab:efficiency}
\small
\begin{tabularx}{\linewidth}{@{}Yrrr@{}}
\toprule
\textbf{Metric} & \textbf{No Skills} & \textbf{S-Full-20} & \textbf{Reduction} \\
\midrule
Input + Output Tokens & 50,025,849 & 40,192,203 & 19.66\% \\
Non-Cached Input Tokens & 3,211,135 & 612,849 & 80.91\% \\
Completed Dev Turns & 27.00 & 26.50 & 1.85\% \\
\textbf{Coding Time (Mins)} & \textbf{189.00} & \textbf{84.46} & \textbf{55.3\%} \\
\bottomrule
\end{tabularx}
\end{table}

Table~\ref{tab:efficiency} reveals a massive 55.3\% reduction in total coding time (saving over 104 minutes per run). Crucially, this time saving is not achieved by simply cutting conversations short (completion turns only decreased by 1.85\%). Furthermore, network logs reveal that reduced external wait times account for only $\sim$6 minutes of this saving. The true source of this acceleration lies in how the skill bank fundamentally alters the model's internal cognitive loop: by navigating the agent away from catastrophic architectural errors early, it slashes non-cached input tokens by a staggering 80.91\%, eliminating redundant context bloat.

\subsection{Ablation Insights: Capability Trumps History Volume}
To disentangle the source of this efficiency, we analyze our two targeted ablations. \textit{S-Ablated-20} tests evidence deprivation (removing 69.66\% of historical text prior to distillation), while \textit{Coreless-3} tests capability deprivation (surgically retaining only 3 basic interface skills).

\begin{table}[htbp]
\centering
\caption{Aggregate Correctness and Coding Time Across All Configurations}
\label{tab:ablation_overview}
\small
\begin{tabularx}{\linewidth}{@{}Yrr@{}}
\toprule
\textbf{Configuration} & \textbf{CCPF} & \textbf{Avg. Coding Time (Mins)} \\
\midrule
No Skills & 46.63\% & 189.00 \\
S-Full-20 & 51.25\% & \textbf{84.46} \\
S-Ablated-20 & \textbf{52.60\%} & 91.74 \\
Coreless-3 & 48.36\% & $\geq$190.87 \\
\bottomrule
\end{tabularx}
\end{table}

\paragraph{History volume is not skill value.} 
Despite a 69.66\% reduction in raw historical text, \textit{S-Ablated-20} surprisingly achieves the highest CCPF (52.60\%), albeit with a slight execution cost penalty (+8.6\% coding time compared to \textit{S-Full-20}). This confirms that as long as the distillation pipeline abstracts the correct engineering principles, effective skill transfer can be sustained by a substantially reduced evidence base. The volume of historical experience is secondary to the quality of the abstracted knowledge.

\paragraph{Execution degradation under Coreless-3.}
Conversely, while \textit{Coreless-3} represents a compound reduction in both skill count and content, its observed execution degradation supports the interpretation that abstracted engineering capabilities contribute to the performance gains of \textit{S-Full-20}. When the targeted core software engineering skills are removed from the supplied memory, the configuration fails to preserve the correctness--time profile. CCPF drops to 48.36\%, and the average coding time skyrockets back to over 190 minutes---worse than the \textit{No Skills} baseline.

\subsection{Behavioral Audit: Decisive Execution vs. Trial-and-Error}
To understand the mechanics behind the \textit{Coreless-3} collapse and the \textit{S-Full-20} acceleration, we audit internal execution actions across all 48 runs (Table~\ref{tab:behavior}).

\begin{table}[htbp]
\centering
\caption{Behavioral Action Audits (Aggregate Totals Across 12 Runs)}
\label{tab:behavior}
\small
\begin{tabularx}{\linewidth}{@{}Yrrrr@{}}
\toprule
\textbf{Action / Event} & \textbf{No Skills} & \textbf{S-Full-20} & \textbf{S-Ablated-20} & \textbf{Coreless-3} \\
\midrule
Tool Invocations & 6,484 & 4,850 & 5,979 & 8,195 \\
Return Text (Chars) & 13.66M & 6.06M & 10.11M & 14.76M \\
Patch Invocations & 865 & 607 & 751 & 988 \\
Added Patch Lines & 40,393 & 38,914 & 41,500 & 46,059 \\
Reasoning Tokens & 441,803 & 302,016 & 360,958 & 699,961 \\
Context Compactions & 21 & 11 & 16 & 26 \\
\bottomrule
\end{tabularx}
\end{table}

The behavioral data reveals two distinct execution paradigms:

\begin{enumerate}
    \item \textbf{Concentrated Edits over Fragmented Patches (\textit{S-Full-20}):} Guided by the complete skill bank, the agent invoked the patch tool 29.8\% less often than the baseline, yet total lines added only decreased by 3.66\%. This means edits grew from an average of 47 to 64 lines per patch. The skill bank helps the agent consolidate its architectural plans, acting with decisive execution rather than falling into a fragmented ``small edit $\rightarrow$ check $\rightarrow$ small edit'' loop.
    \item \textbf{Token-Heavy Thrashing (\textit{Coreless-3}):} Without core capabilities, the agent's execution profile becomes severely bloated. Compared to \textit{S-Full-20}, tool invocations spike by 68.97\%, and generated reasoning tokens explode by 131.76\% (reaching nearly 700k tokens). Without the execution compass, the agent relies on exhaustive, blind exploration, flooding the context window with 14.76M characters of error logs and forcing 26 context compactions.
\end{enumerate}

Ultimately, EngramBench makes these dynamics visible. \textit{Fewer skills do not necessarily mean less work; rather, the absence of abstracted core capabilities forces the base model into catastrophic, token-heavy trial-and-error.}

\section{Limitations and Future Work}
\label{sec:limitations}

While EngramBench establishes a rigorous framework for evaluating skill evolution, we acknowledge two boundary conditions that direct future research:

\paragraph{Computational Scale.} 
Authentic long-horizon execution is computationally exorbitant. Because a single run can consume up to 66 million tokens and require manual trajectory auditing, our deep ablation study is concentrated on 48 strictly controlled runs across 4 representative tasks. Scaling this high-resolution audit to the entire 13-task suite remains an ongoing effort.

\paragraph{Evaluation Scope and Infrastructure.} 
To causally isolate the algorithm's \textit{distillation} capability, we evaluated a strictly frozen skill bank, omitting the complexities of dynamic memory retrieval (e.g., Active RAG). Furthermore, our empirical findings are currently grounded in GPT-5.5 Medium. Future work will test diverse open-weights architectures and enforce stricter, air-gapped container boundaries to further immunize the hidden evaluators.

\section{Conclusion}
\label{sec:conclusion}

In this work, we exposed a critical vulnerability in how the community evaluates skill evolution in autonomous agents: the structural confounding of genuine capability abstraction with rote solution leakage. To resolve this identifiability problem, we introduced EngramBench, the first capability-grounded benchmark governed by the strict axiom of \textit{capability overlap without solution overlap}. 

Crucially, EngramBench shifts the evaluation focus beyond mere functional correctness to equally prioritize execution efficiency—strictly measuring token consumption, development time, and the elimination of meaningless code churn. Our rigorous empirical analysis across 48 multi-hour execution trajectories—further corroborated by human-expert validation to confirm the intrinsic utility of the tasks and distilled skills—revealed a profound insight into procedural memory. We demonstrated that static skill banks do not magically bypass the ``last mile'' of exact code implementation, which remains bottlenecked by the base model's inherent reasoning limits. However, they serve as an indispensable execution compass. By navigating agents away from catastrophic, token-heavy trial-and-error and redundant code modifications, genuine capability abstraction drastically streamlines the internal cognitive loop—slashing context bloat and reducing overall coding time by over 55\%.

Ultimately, EngramBench shifts the agent evaluation paradigm from trivial pattern matching to the verification of deep, cross-domain capability transfer. We provide this infrastructure to the community as a definitive standard, ensuring that future advancements in lifelong learning agents are measured by their true ability to abstract, generalize, and engineer efficiently.

\printbibliography

\clearpage
\appendix
\section{Learning Evidence Coverage Across the Transfer Set}
\label{app:evidence-mapping}

Table~\ref{tab:evidence-mapping-full} extends the representative mappings in
Section~\ref{sec:construction} to all 13 transfer tasks, followed by a shared
cross-layer acceptance mapping. Each row links a public requirement to a reusable
engineering mechanism, a recorded learning episode, and the corresponding
evaluation dimension. Coverage is complete at the task level, not an exhaustive
audit of every README clause; a mechanism-level link does not establish that a
skill bank extracted the mechanism or that an agent used it successfully.

\noindent\textbf{Evidence notation.}
\textit{Repair loop} denotes an observed failure or incorrect result, an applied
change, and subsequent executed verification. \textit{Implementation + verification}
denotes implemented behavior followed by executed checks without a recorded
pre-repair failure of that same behavior. \textit{HTTP observation} distinguishes
observed responses from assertion-based verification. T denotes the turn number
in the archived learning trajectory: all cited turns are from V2 migration
histories except ArtifactVault T26, which is from its original development
history. Case IDs refer to the evaluator version audited on 30 September 2026,
not to historical pass results.

\begingroup
\setcounter{table}{0}
\renewcommand{\thetable}{\thesection.\arabic{table}}
\fontsize{9.2}{11.4}\selectfont
\setlength{\tabcolsep}{3.5pt}
\setlength{\LTleft}{0pt plus 1fill}
\setlength{\LTright}{0pt plus 1fill}
\setlength{\LTcapwidth}{\linewidth}
\renewcommand{\arraystretch}{1.10}
\begin{longtable}{@{}>{\raggedright\arraybackslash}p{0.19\linewidth}>{\raggedright\arraybackslash}p{0.15\linewidth}>{\raggedright\arraybackslash}p{0.13\linewidth}>{\raggedright\arraybackslash}p{0.33\linewidth}>{\raggedright\arraybackslash}p{\dimexpr0.20\linewidth-8\tabcolsep\relax}@{}}
\caption{Requirement-level capability mapping across all 13 transfer tasks and a shared cross-layer acceptance mapping.}
\label{tab:evidence-mapping-full}\\
\toprule
\textbf{Transfer task and public requirement} &
\textbf{Reusable mechanism} &
\textbf{Learning sources} &
\textbf{Recorded failure, repair, and verification evidence} &
\textbf{Evaluation dimension / cases} \\
\midrule
\endfirsthead
\multicolumn{5}{@{}l@{}}{\small\textbf{Table~\thetable\ (continued)}}\\[4pt]
\toprule
\textbf{Transfer task and public requirement} &
\textbf{Reusable mechanism} &
\textbf{Learning sources} &
\textbf{Recorded failure, repair, and verification evidence} &
\textbf{Evaluation dimension / cases} \\
\midrule
\endhead
\midrule
\multicolumn{5}{r@{}}{\footnotesize Continued on next page}\\
\endfoot
\bottomrule
\endlastfoot

\textbf{MeterSettle}\par
Select rates and billing periods by event time; append corrections without
destroying historical records or their read relationships.
& Event-time version binding; append-only correction; consistent read projections.
& GeoPulse T15; BillForge T7.
& \textit{Implementation + verification:} GeoPulse bound historical queries and
events to the applicable revision and ran integration regression checks.
\textit{Repair loop:} BillForge's adjustment returned HTTP 500; a migration fix
enabled creation, but ledger reads still omitted the adjustment. A query repair
restored PAYMENT and ADJUSTMENT entries in subsequent HTTP output.
These episodes do not establish MeterSettle's specific pricing arithmetic.
& \textbf{B}: rating, correction, concurrency.\par
B-01; B-07.\\
\addlinespace[7pt]

\textbf{DockChain}\par
Allocate Berth, Tug, and Yard resources as a complete bundle; two linked Movements
must commit atomically without exceeding capacity.
& Multi-resource commit boundaries; capacity contention; transactional idempotency.
& QuotaMesh T6; BillForge T12.
& \textit{Implementation + verification:} QuotaMesh repaired allocation/revision
boundaries and ran integration tests including hierarchical capacity contention.
BillForge repaired nested database connections, shared the transaction between
business writes and idempotency, and executed concurrent same-key checks with
a single-record assertion. The cited history does not contain a complete
multi-resource deadlock-to-lock-order repair loop.
& \textbf{B}: bundle allocation, concurrency.\par
B-02; B-09.\\
\addlinespace[7pt]

\textbf{IncidentRelay}\par
After an unknown Domain Event ACK, retry the same event identity/body while
preserving per-aggregate order.
& Durable outbox; unknown-result retry; stable event identity; ordered delivery.
& QueueForge T13; ConfigRelay T12.
& \textit{Implementation + fault injection:} QueueForge killed the dispatcher
after webhook 2xx but before ACK persistence, restarted it, and checked
redelivery of the same event ID. ConfigRelay used an HTTP receiver returning
503 on the first request, then asserted stable event identity and that later
events did not overtake the unacknowledged event. These were controlled
fault-injection checks, not observed production incidents.
& \textbf{C}: delivery, recovery.\par
C-07: unknown webhook ACK.\\
\addlinespace[7pt]

\textbf{FlagFoundry}\par
Advance progressive rollout over frozen cohorts; resolve pass/rollback at the
deadline without letting late Outcomes rewrite the terminal state.
& Frozen target sets; staged progression; deadline-triggered durable compensation.
& ConfigRelay T6--T7.
& \textit{Implementation + verification:} ConfigRelay implemented frozen cohorts
and automatic rollback. Executed HTTP/worker scenarios observed a successful
first cohort, a failed second cohort, completed ROLLBACK Work, and delivered
rollback commands. This supports the progression and compensation mechanisms,
not verification of FlagFoundry's bucket algorithm or all Outcome/deadline races.
& \textbf{B}: rollout correctness.\par
B-09: Outcome/deadline race.\\
\addlinespace[7pt]

\textbf{CarbonLedger}\par
Expose a Certificate only with the RETIRED commit; recover after file generation
without duplicate certificates or deductions.
& File/DB publication consistency; recoverable intermediate state; lease fencing.
& ArtifactVault original T26; MediaDock T8.
& \textit{Repair loop:} ArtifactVault's 81\,ms retry interval failed its recovery
check; worker and test-isolation changes were followed by a passing re-test.
\textit{Implementation + verification:} the same turn added recovery from a
renamed blob before DB commit and checked a unique commit after SIGKILL.
MediaDock asserted publication CAS, revision-bound grants, and downloaded bytes.
No pre-fix execution failure was found for the rename/DB gap itself.
& \textbf{C}: worker recovery, persistence.\par
C-03: post-generation SIGKILL.\\
\addlinespace[7pt]

\textbf{ParcelFlow}\par
Split an Order into warehouse Fulfillments; allow at most one Shipment per group
and keep the parent terminal state consistent with every group.
& Frozen membership; fan-out; child/parent state consistency; atomic persistence.
& ConfigRelay T6--T7; BillForge T12.
& \textit{Implementation + verification:} ConfigRelay's frozen cohorts,
per-member progression, and asynchronous rollback produced actual HTTP and
Work results. BillForge supplied repaired transaction boundaries and executed
idempotency regression checks. These are composable mechanisms, not evidence
that the learning tasks already verified cross-warehouse greedy allocation or
Shipment/cancel business rules.
& \textbf{B}: grouped allocation, concurrency.\par
B-03; B-10.\\
\addlinespace[7pt]

\textbf{ColdChain\-Control}\par
Late readings may correct excursion history without regressing the current
projection; Recall freezes its target set at publication.
& Event-time rules; late-event classification; history/current separation;
frozen membership.
& GeoPulse T14--T15; ConfigRelay T6--T7.
& \textit{Implementation + verification:} GeoPulse asserted late-event
classification and continuous transition sequences in T14, then repaired
historical revision selection and ran regression checks in T15.
ConfigRelay's staged cohorts supplied executed frozen-membership behavior.
These excerpts do not directly establish a complete excursion rebuild or
non-regression of the current route.
& \textbf{B}: projection, concurrency.\par
B-06; B-10.\\
\addlinespace[7pt]

\textbf{CreatorRights\-Exchange}\par
Correct CLOSED royalties only by appending to a later OPEN period; preserve
old postings/periods and trace each adjustment.
& Append-only ledger correction; compensation rather than rewriting;
original-entry links and read projections.
& BillForge T7; LedgerBridge T6.
& \textit{Repair loop:} BillForge moved from adjustment HTTP 500 to successful
creation with missing ledger visibility, then repaired the query and observed
both entry types. \textit{Implementation + HTTP observation:} LedgerBridge added
REVERSAL; reversal and same-key replay matched, and both statements retained
TRANSFER/REVERSAL entries. The single-leg scenario is not complete cross-period
financial verification.
& \textbf{B}: data correctness, idempotency, concurrency.\par
B-10: adjustment uniqueness/balance.\\
\addlinespace[7pt]

\textbf{AccessSentinel}\par
After revocation commits, checks must fail closed; stale credentials/caches
cannot authorize, and concurrent operations must use current authority.
& Credential generations; rotation/reuse; monotonic revocation;
conditional checks on authoritative state.
& IdentityMesh T10/T12; ClinicGrid T43.
& \textit{Implementation + HTTP observation:} IdentityMesh recorded refresh
replay/reuse, revoked-subject rejection, and stale key-rotation rejection.
\textit{Repair loop:} after a Work state/terminal HTTP 500, ClinicGrid changed
to a DB-conditional claim; the re-test no longer produced that error.
The latter supports atomic authority checks, not complete multi-API cached
revocation behavior in AccessSentinel.
& \textbf{B}: authority, idempotency, concurrency.\par
B-09: cross-process revocation.\\
\addlinespace[7pt]

\textbf{Commerce\-Command}\par
Retain UNKNOWN after a provider accepts payment but loses the response;
recover with stable request identity without another charge or journal entry.
& Unknown versus failed outcomes; stable side-effect identity; durable retry;
transactional records.
& QueueForge T13; BillForge T12.
& \textit{Implementation + fault injection:} QueueForge's kill-before-persisted-ACK
check exercised a side effect whose response had already arrived, followed by
redelivery after restart. BillForge supplied repaired joint business/idempotency
commit behavior and concurrent same-key regression checks. This is a
mechanism-level composition; webhook redelivery does not itself verify
exactly-once payment charging.
& \textbf{C}: worker recovery, persistence.\par
C-05: payment reconciliation.\\
\addlinespace[7pt]

\textbf{EscrowGuard}\par
Conserve available/released/refunded funds; create all Beneficiary Payouts
atomically, without releasing and refunding the same amount twice.
& Integer fund conservation; immutable compensation entries;
same-transaction idempotency and settlement.
& LedgerBridge T6; BillForge T12.
& \textit{Implementation + HTTP observation:} LedgerBridge appended reversal
entries and observed both parties' statements and replay.
\textit{Repair + verification:} BillForge repaired shared transaction handling
and ran concurrent idempotency checks. These episodes supply financial
correction and commit-boundary experience; the single-leg history does not
establish complete multi-beneficiary contention behavior.
& \textbf{B}: fund correctness, concurrency.\par
B-01; B-09.\\
\addlinespace[7pt]

\textbf{PermitForge}\par
Only a valid current-Stage Claim may write a Decision; stage evidence and
revisions must remain correctly scoped.
& Stage-local identity; Claim/revision fencing; decision-to-audit consistency.
& Moderation\-Flow T3--T4; ClinicGrid T43.
& \textit{Implementation + assertion-based verification:} ModerationFlow
implemented revision/lease-aware claims and exercised real HTTP claim-to-decision
flows with Case, audit, and snapshot assertions. ClinicGrid supplied its
DB-conditional claim repair. These positive flows do not verify every stale
Claim, role quorum, or repeated reviewer across Stages. A-14 reflects the current
public Stage-policy revision, not an unchanged V1 requirement.
& \textbf{A}: public contract, A-14.\par
\textbf{B}: concurrency, B-10.\\
\addlinespace[7pt]

\textbf{CapacityLease}\par
Preserve capacity and atomic Gang state; recover expired leases, fence stale
owners, and actually drain the backlog.
& Capacity/commit invariants; DB leases and fencing; recovery throughput;
post-condition checks.
& QuotaMesh T6; QueueForge T16.
& \textit{Implementation + verification:} QuotaMesh repaired capacity/revision
boundaries and ran hierarchical contention tests. \textit{Repair loop:}
QueueForge recovered only 572/2,000 expired leases; batch-result commits and
SKIP LOCKED changes raised recovery to 2,000/2,000. Its test query also needed
a join-duplicate counting fix before the final public performance command
passed. This is not proof of complete Gang fairness.
& \textbf{B}: B-09, B-10.\par
\textbf{C}: C-05, expired-lease fencing.\par
\textbf{E}: E-06, recovery at scale.\\
\addlinespace[7pt]
\midrule

\textbf{Shared cross-layer acceptance}\par
CapacityLease and CreatorRights\-Exchange: production UI actions must traverse
real APIs and appear in persistent state/public projections.
& Browser-to-API-to-storage closure; independent acceptance;
non-empty projection checks.
& MergeBoard T5; Evidence\-Chain T10; MediaDock T8.
& \textit{Repair loops:} MergeBoard's Chromium test could not find controls;
production asset-loading repairs were followed by a passing browser test.
EvidenceChain's split succeeded but snapshot returned HTTP 500; adding the
parent's aliquots repaired the subsequent HTTP chain.
\textit{Implementation + verification:} MediaDock compared actual downloaded
bytes with uploaded bytes. These are concrete exercised paths, not proof of
whole-project acceptance.
& \textbf{D}: OpenAPI, UI, cross-layer checks.\par
CapacityLease: D-01, D-02, D-08.\par
CreatorRights\-Exchange: D-02, D-08.\\

\end{longtable}
\endgroup

\end{document}